\documentclass[11pt]{article}
\usepackage{amsmath,latexsym,amssymb}

\usepackage{hyperref}
\usepackage{tikz}
\usetikzlibrary{arrows,calc}
\usepackage{pdflscape}
\usepackage{rotating}
\usepackage{graphics}
\usepackage{enumerate}
\newif\ifpdf
\pdffalse
\pdftrue
\ifpdf

\else

\fi
\newcommand{\ignore}[1]{}
\newcommand{\mfalls}{\quad\mbox{if \;}}
\renewcommand{\cases}[1]{\left\{\begin{array}{rl}#1\end{array}\right.}

\newcommand{\E}{\mathbf{E}}

\newcommand{\R}{{\mathbb{R}}}

\newcommand{\bec}{\begin{equation}}
\newcommand{\eec}{\end{equation}}
\newcommand{\bac}{\begin{eqnarray}}
\newcommand{\eac}{\end{eqnarray}}
\newcommand{\be}{\begin{displaymath}}
\newcommand{\ee}{\end{displaymath}}
\newcommand{\ba}{\begin{eqnarray*}}
\newcommand{\ea}{\end{eqnarray*}}

\newtheorem{proposition}{Proposition}[section]

\newtheorem{uremark}[proposition]{Remark}

\newtheorem{uexample}[proposition]{Example}

\def\qed{\mbox{$\Box$}}

\newtheorem{udefi}{Definition}[section]
\newtheorem{utheo}{Theorem}
\newtheorem{usatz}[udefi]{Satz}
\newtheorem{uprop}[udefi]{Proposition}

\newtheorem{ubemerkung}[udefi]{Bemerkung}
\newtheorem{ukorollar}[udefi]{Korollar}

\usepackage{color}
\definecolor{Red}{rgb}{1,0,0}

\newcommand{\msonst}{\quad\mbox{otherwise}}
 
\begin{document}
\title{Mean Number of Visible Confetti\\}
\author{Achim Klenke
\\
Johannes Gutenberg-Universit{\"a}t Mainz\\Institut f{\"u}r Mathematik\\
Staudingerweg 9\\ 55099 Mainz\\ Germany\\
math@aklenke.de}
\date{\small 14.12.2022}\maketitle
\begin{abstract}{\small We use a Monte Carlo simulation to estimate the mean number of visible confetti per unit square when the confetti are placed successively at random positions.}
\end{abstract}
{\noindent AMS 2000 subject classification: 60D05 65C05.}
\section{Introduction}
\subsection{Motivation and Main result}
Consider confetti of a fixed radius $r$ thrown at random on a glass floor until the floor is covered. How many confetti per square meter are visible (on average) \emph{from below}? This is a special case of the so-called dead leaves model introduced by Matheron in \cite{Matheron1968,Matheron1975}. Note that the picture from below is the same (in distribution) as the picture from above if we observe the stationary distribution of the Markov chain with values in the confetti configurations that is defined by throwing the confetti one by one. In this sense, it is an example for the perfect simulation scheme \emph{coupling from the past}. See, e.g., \cite{KendallThoennes1999}.

The answer seems to be unknown although the following seemingly more complicated question has been answered: If we count the different faces that are visible from below, then on average we have $\frac4\pi r^{-2}$ faces per unit area. See \cite[Theorem 3.4]{Penrose2020} with $\lambda=\pi r^2$ and $\mathcal{H}_1(\partial S)=2\pi r$. Note that a visible confetto (this seems to be the correct singular) can show more than one face if it is covered partially by two (or more) confetti. See Figure~\ref{F1}. Hence, the average number of visible confetti per unit area should be $c r^{-2}$ with some $c<\frac4\pi\approx1.2732$.

The main purpose of this paper is to present a Monte Carlo simulation to estimate
\begin{equation}
\label{E01}
c=1.146015(33).
\end{equation}
\bigskip
\definecolor{F0}{rgb}{0.9,0.9,0.9}
\definecolor{C30}{rgb}{0.9,0.3,0.3}
\definecolor{C29}{rgb}{0.8,0.90,0.95}
\definecolor{C28}{rgb}{0.95,0.95,0}
\definecolor{C27}{rgb}{0.3,0.9,0.3}
\definecolor{C26}{rgb}{0.9,0.3,0.3}
\definecolor{C25}{rgb}{0.8,0.90,0.95}
\definecolor{C24}{rgb}{0.95,0.95,0}
\definecolor{C23}{rgb}{0.3,0.9,0.3}
\definecolor{C22}{rgb}{0.6,0.6,1}
\definecolor{C21}{rgb}{0.9,0.3,0.3}
\definecolor{C20}{rgb}{0.65,0.0,1}
\definecolor{C19}{rgb}{1.0,0.62,0.0}
\definecolor{C18}{rgb}{0.4,0.6,0.6}
\definecolor{C17}{rgb}{0.8,0.4,0.6}
\definecolor{C16}{rgb}{1.0,0.9,0.4}
\definecolor{C15}{rgb}{0.6,0.8,1.0}
\definecolor{C14}{rgb}{0.6,0.3,0.4}
\definecolor{C13}{rgb}{1.0,1.0,0}
\definecolor{C12}{rgb}{0.8,0.90,0.95}
\definecolor{C11}{rgb}{0.95,0.95,0}
\definecolor{C10}{rgb}{0,1,0}
\definecolor{C09}{rgb}{0.6,0.6,1}
\definecolor{C08}{rgb}{1.0,0.3,0.3}
\definecolor{C07}{rgb}{0.65,0.0,1}
\definecolor{C06}{rgb}{1.0,0.62,0.0}
\definecolor{C05}{rgb}{0.4,0.6,0.6}
\definecolor{C04}{rgb}{0.8,0.4,0.6}
\definecolor{C03}{rgb}{1.0,0.9,0.4}
\definecolor{C02}{rgb}{0.6,0.8,1.0}
\definecolor{C01}{rgb}{0.6,0.3,0.4}
\definecolor{C134}{rgb}{0.6,0.8,0.8}
\begin{figure}[ht]
\centerline{
\begin{tikzpicture}[scale = 0.7]
\def\myrad{3.0}
\def\csize{1.4}
\clip (-3,-3) rectangle +(16,16);
\draw[fill=F0, line width = 0.3mm, shift={(0,0)}] ( 8.95 , 14.103 ) circle(\myrad);
\draw[fill=F0, line width = 0.3mm, shift={(0,0)}] ( 13.678 , 5.67 ) circle(\myrad);
\draw[fill=F0, line width = 0.3mm, shift={(0,0)}] ( 6.843 , -1.5 ) circle(\myrad);
\draw[fill=F0, line width = 0.3mm, shift={(0,0)}] ( 15.691 , 2.793 ) circle(\myrad);
\draw[fill=F0, line width = 0.3mm, shift={(0,0)}] ( 14.413 , 14.203 ) circle(\myrad);
\draw[fill=F0, line width = 0.3mm, shift={(0,0)}] ( 7.851 , 14.928 ) circle(\myrad);
\draw[fill=F0, line width = 0.3mm, shift={(0,0)}] ( -0.006 , -5.232 ) circle(\myrad);
\draw[fill=F0, line width = 0.3mm, shift={(0,0)}] ( 13.148 , 4.691 ) circle(\myrad);
\draw[fill=F0, line width = 0.3mm, shift={(0,0)}] ( 10.407 , -0.479 ) circle(\myrad);
\draw[fill=F0, line width = 0.3mm, shift={(0,0)}] ( -2.922 , -2.868 ) circle(\myrad);
\draw[fill=F0, line width = 0.3mm, shift={(0,0)}] ( 5.19 , -2.705 ) circle(\myrad);
\draw[fill=F0, line width = 0.3mm, shift={(0,0)}] ( 1.225 , -2.718 ) circle(\myrad);
\draw[fill=F0, line width = 0.3mm, shift={(0,0)}] ( 6.259 , -4.702 ) circle(\myrad);
\draw[fill=F0, line width = 0.3mm, shift={(0,0)}] ( 7.828 , 14.878 ) circle(\myrad);
\draw[fill=F0, line width = 0.3mm, shift={(0,0)}] ( 13.438 , 3.435 ) circle(\myrad);
\draw[fill=F0, line width = 0.3mm, shift={(0,0)}] ( 14.853 , 11.337 ) circle(\myrad);
\draw[fill=F0, line width = 0.3mm, shift={(0,0)}] ( -4.415 , 3.931 ) circle(\myrad);
\draw[fill=F0, line width = 0.3mm, shift={(0,0)}] ( -3.045 , 14.03 ) circle(\myrad);
\draw[fill=F0, line width = 0.3mm, shift={(0,0)}] ( -5.287 , -2.287 ) circle(\myrad);
\draw[fill=F0, line width = 0.3mm, shift={(0,0)}] ( 13.698 , 2.302 ) circle(\myrad);
\draw[fill=F0, line width = 0.3mm, shift={(0,0)}] ( 13.714 , 9.917 ) circle(\myrad);
\draw[fill=F0, line width = 0.3mm, shift={(0,0)}] ( -0.819 , 1.429 ) circle(\myrad);
\draw[fill=F0, line width = 0.3mm, shift={(0,0)}] ( 15.911 , 2.058 ) circle(\myrad);
\draw[fill=F0, line width = 0.3mm, shift={(0,0)}] ( 11.244 , -4.582 ) circle(\myrad);
\draw[fill=F0, line width = 0.3mm, shift={(0,0)}] ( -3.826 , 11.064 ) circle(\myrad);
\draw[fill=F0, line width = 0.3mm, shift={(0,0)}] ( 11.789 , 9.838 ) circle(\myrad);
\draw[fill=F0, line width = 0.3mm, shift={(0,0)}] ( -4.718 , 2.844 ) circle(\myrad);
\draw[fill=F0, line width = 0.3mm, shift={(0,0)}] ( -5.362 , -1.219 ) circle(\myrad);
\draw[fill=F0, line width = 0.3mm, shift={(0,0)}] ( 10.741 , 8.5 ) circle(\myrad);
\draw[fill=F0, line width = 0.3mm, shift={(0,0)}] ( 5.598 , 15.703 ) circle(\myrad);
\draw[fill=F0, line width = 0.3mm, shift={(0,0)}] ( -1.28 , -5.495 ) circle(\myrad);
\draw[fill=F0, line width = 0.3mm, shift={(0,0)}] ( 7.401 , 13.494 ) circle(\myrad);
\draw[fill=F0, line width = 0.3mm, shift={(0,0)}] ( 14.391 , 1.316 ) circle(\myrad);
\draw[fill=F0, line width = 0.3mm, shift={(0,0)}] ( 12.334 , -1.036 ) circle(\myrad);
\draw[fill=F0, line width = 0.3mm, shift={(0,0)}] ( 9.574 , -2.943 ) circle(\myrad);
\draw[fill=F0, line width = 0.3mm, shift={(0,0)}] ( 14.833 , -3.483 ) circle(\myrad);
\draw[fill=C12, line width = 0.3mm, shift={(0,0)}] ( 3.412 , 0.324 ) circle(\myrad);
\draw ( 3.7 , -1.9 ) node[anchor=center, scale=\csize, shift={(0,0)}]{$12$};
\draw[fill=C11, line width = 0.3mm, shift={(0,0)}] ( 3.271 , 1.833 ) circle(\myrad);
\draw ( 3.6 , -0.4 ) node[anchor=center, scale=\csize, shift={(0,0)}]{$11$};
\draw[fill=F0, line width = 0.3mm, shift={(0,0)}] ( -3.52 , 7.169 ) circle(\myrad);
\draw[fill=F0, line width = 0.3mm, shift={(0,0)}] ( 12.466 , 13.907 ) circle(\myrad);
\draw[fill=F0, line width = 0.3mm, shift={(0,0)}] ( -0.409 , -2.597 ) circle(\myrad);
\draw[fill=F0, line width = 0.3mm, shift={(0,0)}] ( -2.111 , 7.908 ) circle(\myrad);
\draw[fill=F0, line width = 0.3mm, shift={(0,0)}] ( 0.724 , 15.545 ) circle(\myrad);
\draw[fill=C10, line width = 0.3mm, shift={(0,0)}] ( 8.662 , 1.444 ) circle(\myrad);
\draw ( 7.7 , -0.5 ) node[anchor=center, scale=\csize, shift={(0,0)}]{$10$};
\draw[fill=F0, line width = 0.3mm, shift={(0,0)}] ( -0.026 , -0.054 ) circle(\myrad);
\draw[fill=C09, line width = 0.3mm, shift={(0,0)}] ( 5.792 , 2.829 ) circle(\myrad);
\draw ( 5.1 , 0.9 ) node[anchor=center, scale=\csize, shift={(0,0)}]{$9$};
\draw[fill=F0, line width = 0.3mm, shift={(0,0)}] ( 2.614 , 12.07 ) circle(\myrad);
\draw[fill=F0, line width = 0.3mm, shift={(0,0)}] ( 3.326 , 11.294 ) circle(\myrad);
\draw[fill=F0, line width = 0.3mm, shift={(0,0)}] ( 9.172 , -4.575 ) circle(\myrad);
\draw[fill=F0, line width = 0.3mm, shift={(0,0)}] ( 13.588 , 14.712 ) circle(\myrad);
\draw[fill=F0, line width = 0.3mm, shift={(0,0)}] ( -0.331 , 8.502 ) circle(\myrad);
\draw[fill=F0, line width = 0.3mm, shift={(0,0)}] ( 10.836 , 11.53 ) circle(\myrad);
\draw[fill=C08, line width = 0.3mm, shift={(0,0)}] ( 7.14 , 3.944 ) circle(\myrad);
\draw ( 6.7 , 1.4 ) node[anchor=center, scale=\csize, shift={(0,0)}]{$8$};
\draw[fill=F0, line width = 0.3mm, shift={(0,0)}] ( 13.107 , 12.515 ) circle(\myrad);
\draw[fill=C07, line width = 0.3mm, shift={(0,0)}] ( 0.723 , 1.656 ) circle(\myrad);
\draw ( 0.723 , 1.656 ) node[anchor=center, scale=\csize, shift={(0,0)}]{$7$};
\draw[fill=C06, line width = 0.3mm, shift={(0,0)}] ( 0.71 , 6.503 ) circle(\myrad);
\draw ( 0.5 , 4.303 ) node[anchor=center, scale=\csize, shift={(0,0)}]{$6$};
\draw ( 2.2 , 7.90 ) node[anchor=center, scale=\csize, shift={(0,0)}]{$6$};
\draw[fill=C05, line width = 0.3mm, shift={(0,0)}] ( 4.196 , 4.922 ) circle(\myrad);
\draw ( 2.196 , 4.622 ) node[anchor=center, scale=\csize, shift={(0,0)}]{$5$};
\draw[fill=F0, line width = 0.3mm, shift={(0,0)}] ( 14.8 , -0.937 ) circle(\myrad);
\draw[fill=F0, line width = 0.3mm, shift={(0,0)}] ( -1.18 , 7.33 ) circle(\myrad);
\draw[fill=F0, line width = 0.3mm, shift={(0,0)}] ( -1.036 , 9.787 ) circle(\myrad);
\draw[fill=F0, line width = 0.3mm, shift={(0,0)}] ( 6.868 , 11.216 ) circle(\myrad);
\draw[fill=C04, line width = 0.3mm, shift={(0,0)}] ( 9.933 , 2.244 ) circle(\myrad);
\draw ( 9.033 , 1.244 ) node[anchor=center, scale=\csize, shift={(0,0)}]{$4$};
\draw[fill=F0, line width = 0.3mm, shift={(0,0)}] ( 10.955 , 4.977 ) circle(\myrad);
\draw[fill=C03, line width = 0.3mm, shift={(0,0)}] ( 5.672 , 7.543 ) circle(\myrad);
\draw ( 3.9 , 8.4 ) node[anchor=center, scale=\csize, shift={(0,0)}]{$3$};
\draw[fill=C02, line width = 0.3mm, shift={(0,0)}] ( 7.394 , 7.621 ) circle(\myrad);
\draw ( 7.394 , 8.3 ) node[anchor=center, scale=\csize, shift={(0,0)}]{$2$};
\draw[fill=C01, line width = 0.3mm, shift={(0,0)}] ( 6.028 , 4.713 ) circle(\myrad);
\draw (6.028,4.713) node[anchor=center, scale=\csize, shift={(0,0)}]{$1$};
\draw[fill=F0, line width = 0.3mm, shift={(0,0)}] ( -0.709 , -2.187 ) circle(\myrad);
\draw ( 0.5 , 4.303 ) node[anchor=center, scale=\csize, shift={(0,0)}]{$6$};
\draw ( 2.2 , 7.90 ) node[anchor=center, scale=\csize, shift={(0,0)}]{$6$};

\draw[color = black, fill=none, draw=black, line width = 0.3mm, shift={(0,0)}] (0,0) rectangle (10.0,10.0);
\end{tikzpicture}
}
\caption{Confetti of radius $r=0.4$. The 12 visible confetti with centers in the unit square have colors. They came in the order indicated by the numbers. Note that confetto 6 shows two faces and would be counted twice in the model considered in \cite{Penrose2020}. Also note that confetto 12 has its center in the unit square but is visible outside only; it is still counted in our model.}
\label{F1}
\end{figure}
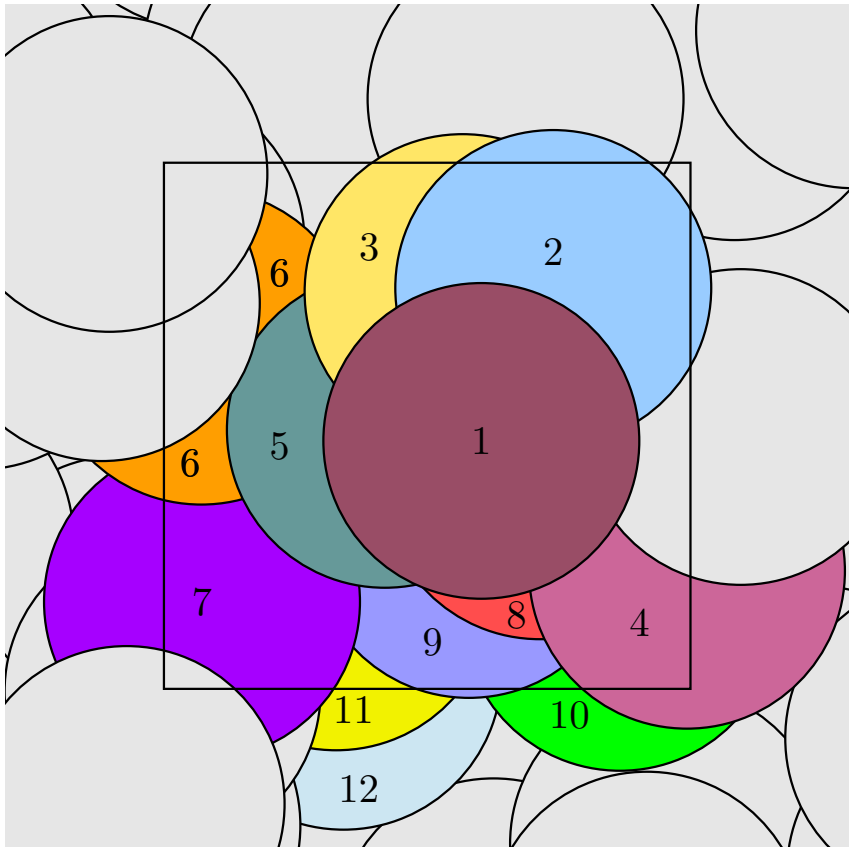

\subsection{Formal description of the model}
In order to define the model formally, let $X$ be a Poisson point process on $\R^2\times[0,\infty)$. Let
$$B_r(x):=\big\{y\in\R^2:\,\|x-y\|<r\big\},\qquad x\in\R^2,$$
be the open disc with radius $r$ centered at $x$.
We consider a point $(x,t)$ of $X$ as a confetto centered at $x$ and thrown by time $t$. We say that $(x,t)$ is visible if the corresponding confetto $B_r(x)$ is visible from below. Formally, for $t>0$, define the set of points covered before time $t$
\begin{equation}
\label{E02}
S_r(t):=\big\{x\in\R^2:\,X\big(B_r(x)\times[0,t)\big)>0\big\}.
\end{equation}
Furthermore, for $(x,t)\in\R^2\times(0,\infty)$, define the function $v_r$ by
\begin{equation}
\label{E03}
v_r((x,t))=\cases{1,&\mfalls B_r(x)\setminus S_r(t)\neq\emptyset,\\[2mm]
0,&\msonst.}
\end{equation}
In words, $v_r(x,t)$ equals $0$ if the confetto centered at $x$ is completely hidden by confetti thrown before time $t$ and is $1$ otherwise. So it is the indicator function for the confetti centers $x$ that are not yet completely covered before time $t$.

Now consider the point process $V$ on $\R^2$ defined by
\begin{equation}
\label{E04}
V_r(A)=\int_{A\times[0,\infty)}X(d(x,t))v_r(x,t),\qquad A\subset\R^2\mbox{ Borel.}
\end{equation}
That is, $V_r(A)$ is the number of visible points from $X$ in the set $A$.
Clearly, $V_r$ is a stationary point process with a finite intensity $c_r=\E\big[V_r\big([0,1]^2\big)\big]$ and $c_r$ fulfills the scaling relation $c_r = c_1 r^{-2}$. Hence, the constant
\begin{equation}
\label{E05}
c:=\E\big[V_1\big([0,1]^2\big)\big]
\end{equation}
is the average number of visible radius 1 confetti per unit area and our simulations show that $c=1.146015(33)$.

\subsection{Simulation}
For the simulation, we consider $V_r\big([0,1]^2\big)$. The simulations are not equally efficient for all values of $r$ and so we do not define $r$ at this point but will choose it later.

For the simulation of $V_r$, we have to consider the larger square $\verb"S2":=[-2r,1+2r]^2$. In fact, a point in $\verb"S0":=[0,1]^2$ can be visible since the corresponding confetto is visible in $\verb"S1":=[-r,1+r]^2$. In order to check visibility in this larger square, we need to consider all points of $X$ in the even larger square $S2$.

The na\"{\i}ve algorithm to generate a sample of $V_r\big([0,1]^2\big)$ now works as follows.
\begin{verbatim}
Phase 1 (naive).
List_of_points = empty
REPEAT
  Place a point P at a random position in S2
  Add P to List_of_points
UNTIL every point in S1 is covered
\end{verbatim}
\clearpage
\begin{verbatim}
Phase 2.
Counter = 0
For every P from List_of_points
  if P is in S0 and P is visible
    Counter = Counter + 1
Return Counter
\end{verbatim}

The time-consuming parts are checking if \verb"S1" is covered in Phase 1 and checking visibility of \verb"P" in Phase 2. Hence, it turns out that a doubling scheme in Phase 1 is more efficient than adding points one by one. Also, it turns out that we gain efficiency by sorting out the invisible points already in Phase 1. Unfortunately, due to numerical errors, in very few cases our algorithm sorted out all new points as invisible although \verb"S1" was not covered completely. We helped this issue by sorting out the invisible points for the smaller radius $0.99\cdot r$ instead of $r$. As the pre-sorting only helps with efficiency and since we count the visible points for the correct radius $r$ later in Phase 2, this does not change the results that we would get without the pre-sorting.
To conclude, instead of Phase 1 (na\"{\i}ve) above, we do the following:
\begin{verbatim}
Phase 1 (more efficient).
List_of_points = empty
n = 1
REPEAT
  REPEAT n times
    Place a point P at a random position in S2
    If P is not in S1 or P is visible for radius 0.99*r
      Add P to List_of_points
  n = 2 * n
UNTIL every point in S1 is covered
\end{verbatim}
\definecolor{C21}{rgb}{0.9,0.3,0.3}
\definecolor{C08}{rgb}{0.9,0.3,0.3}
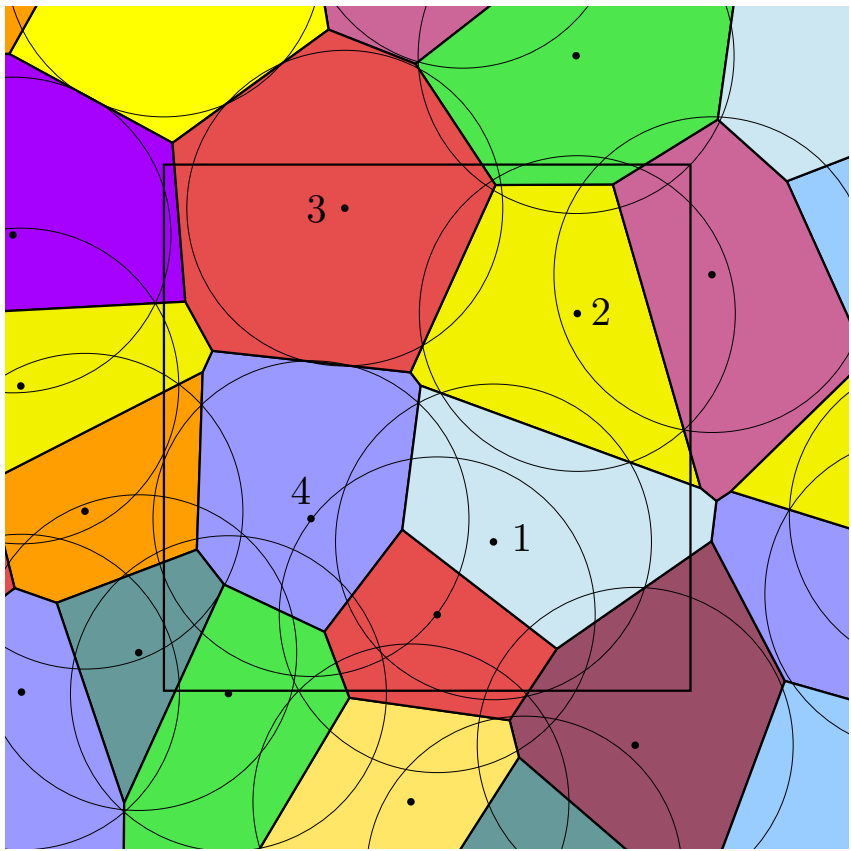
\begin{figure}[ht]
\centerline{
\begin{tikzpicture}[scale = 0.7]
\def\myrad{3.0}
\def\kleinrad{0.05}
\def\csize{1.4}
\clip (-3,-3) rectangle +(16,16);
\draw[fill=C01, line width =0.3mm] (11.79339999,0.185420022) -- (10.398389984,2.84174) -- (7.460729986,0.796400002) -- (6.565069974,-0.56153) -- (6.742939978,-1.266219986) -- (10.159659986,-4.208199988) --  cycle;
\draw[fill=C02, line width =0.3mm] (18,6.885319982) -- (18,10.025979998) -- (14.958959998,10.893489992) -- (11.830039996,9.68375999) -- (13.351379998,6.30869999) --  cycle;
\draw[fill=C03, line width =0.3mm] (15.500369984,-1.812209976) -- (14.457689982,-1.169749976) -- (11.899189986,-5.088849992) --  cycle;
\draw[fill=C04, line width =0.3mm] (11.121519976,16.866589982) -- (11.411299986,18) -- (2.239790002,18) -- (3.126390002,12.565539982) -- (4.779180024,11.91299999) --  cycle;
\draw[fill=C05, line width =0.3mm] (10.507019986,-6.680109974) -- (10.159659986,-4.208199988) -- (6.742939978,-1.266219986) -- (2.449850008,-8) -- (10.500389986,-8) --  cycle;
\draw[fill=C06, line width =0.3mm] (0.624810012,2.681170006) -- (0.735720006,6.057819984) -- (-3.34000999,3.972020008) -- (-2.836109996,1.950130008) -- (-2.037109976,1.670080004) --  cycle;
\draw[fill=C07, line width =0.3mm] (18,2.560410016) -- (16.23848999,2.083360014) -- (13.441729998,-0.28525998) -- (14.457689982,-1.169749976) -- (15.500369984,-1.812209976) -- (18,-2.779809986) --  cycle;
\draw[fill=C22, line width =0.3mm] (4.52734002,3.060520016) -- (4.876450002,5.798129982) -- (4.681620016,6.050949978) -- (0.919790022,6.460219982) -- (0.735720006,6.057819984) -- (0.624810012,2.681170006) -- (1.141280018,2.017700004) -- (3.047380006,1.118820022) --  cycle;
\draw[fill=C09, line width =0.3mm] (16.23848999,2.083360014) -- (10.767419996,3.787340006) -- (10.491679986,3.60551002) -- (10.398389984,2.84174) -- (11.79339999,0.185420022) -- (13.441729998,-0.28525998) --  cycle;
\draw[fill=C10, line width =0.3mm] (18,-2.779809986) -- (15.500369984,-1.812209976) -- (11.899189986,-5.088849992) -- (10.507019986,-6.680109974) -- (10.500389986,-8) -- (18,-8) --  cycle;
\draw[fill=C11, line width =0.3mm] (8.526399994,9.622719998) -- (6.296949992,9.61225999) -- (4.681620016,6.050949978) -- (4.876450002,5.798129982) -- (10.196389978,3.846670004) --  cycle;
\draw[fill=C12, line width =0.3mm] (18,18) -- (11.411299986,18) -- (11.121519976,16.866589982) -- (11.18722,15.688629978) -- (14.958959998,10.893489992) -- (18,10.025979998) --  cycle;
\draw[fill=C13, line width =0.3mm] (3.126390002,12.565539982) -- (2.239790002,18) -- (0.375100006,18) -- (-2.937319988,12.103169996) -- (0.162560016,10.411679988) --  cycle;
\draw[fill=C14, line width =0.3mm] (-4.146769996,0.95869) -- (-8,1.634250002) -- (-8,-8) -- (-2.788959984,-8) --  cycle;
\draw[fill=C15, line width =0.3mm] (11.899189986,-5.088849992) -- (14.457689982,-1.169749976) -- (13.441729998,-0.28525998) -- (11.79339999,0.185420022) -- (10.159659986,-4.208199988) -- (10.507019986,-6.680109974) --  cycle;
\draw[fill=C16, line width =0.3mm] (6.742939978,-1.266219986) -- (6.565069974,-0.56153) -- (3.520780024,-0.13422) -- (-0.792449988,-7.401929982) -- (-0.961049978,-8) -- (2.449850008,-8) --  cycle;
\draw[fill=C17, line width =0.3mm] (10.767419996,3.787340006) -- (13.351379998,6.30869999) -- (11.830039996,9.68375999) -- (10.518939998,10.85741) -- (8.526399994,9.622719998) -- (10.196389978,3.846670004) -- (10.491679986,3.60551002) --  cycle;
\draw[fill=C18, line width =0.3mm] (1.141280018,2.017700004) -- (0.624810012,2.681170006) -- (-2.037109976,1.670080004) -- (-0.754929986,-2.13542999) --  cycle;
\draw[fill=C19, line width =0.3mm] (-2.937319988,12.103169996) -- (0.375100006,18) -- (-8,18) -- (-8,12.063449978) --  cycle;
\draw[fill=C20, line width =0.3mm] (0.40671,7.394229994) -- (0.162560016,10.411679988) -- (-2.937319988,12.103169996) -- (-8,12.063449978) -- (-8,6.954859998) --  cycle;
\draw[fill=C21, line width =0.3mm] (7.460729986,0.796400002) -- (4.52734002,3.060520016) -- (3.047380006,1.118820022) -- (3.520780024,-0.13422) -- (6.565069974,-0.56153) --  cycle;
\draw[fill=C22, line width =0.3mm] (-0.792449988,-7.401929982) -- (-0.754929986,-2.13542999) -- (-2.037109976,1.670080004) -- (-2.836109996,1.950130008) -- (-4.146769996,0.95869) -- (-2.788959984,-8) -- (-0.961049978,-8) --  cycle;
\draw[fill=C23, line width =0.3mm] (3.520780024,-0.13422) -- (3.047380006,1.118820022) -- (1.141280018,2.017700004) -- (-0.754929986,-2.13542999) -- (-0.792449988,-7.401929982) --  cycle;
\draw[fill=C24, line width =0.3mm] (0.735720006,6.057819984) -- (0.919790022,6.460219982) -- (0.40671,7.394229994) -- (-8,6.954859998) -- (-8,6.88120998) -- (-3.34000999,3.972020008) --  cycle;
\draw[fill=C25, line width =0.3mm] (10.398389984,2.84174) -- (10.491679986,3.60551002) -- (10.196389978,3.846670004) -- (4.876450002,5.798129982) -- (4.52734002,3.060520016) -- (7.460729986,0.796400002) --  cycle;
\draw[fill=C26, line width =0.3mm] (-2.836109996,1.950130008) -- (-3.34000999,3.972020008) -- (-8,6.88120998) -- (-8,1.634250002) -- (-4.146769996,0.95869) --  cycle;
\draw[fill=C27, line width =0.3mm] (8.526399994,9.622719998) -- (10.518939998,10.85741) -- (11.18722,15.688629978) -- (11.121519976,16.866589982) -- (4.779180024,11.91299999) -- (6.296949992,9.61225999) --  cycle;
\draw[fill=C28, line width =0.3mm] (18,2.560410016) -- (18,6.885319982) -- (13.351379998,6.30869999) -- (10.767419996,3.787340006) -- (16.23848999,2.083360014) --  cycle;
\draw[fill=C29, line width =0.3mm] (14.958959998,10.893489992) -- (11.18722,15.688629978) -- (10.518939998,10.85741) -- (11.830039996,9.68375999) --  cycle;
\draw[fill=C30, line width =0.3mm] (6.296949992,9.61225999) -- (4.779180024,11.91299999) -- (3.126390002,12.565539982) -- (0.162560016,10.411679988) -- (0.40671,7.394229994) -- (0.919790022,6.460219982) -- (4.681620016,6.050949978) --  cycle;
\draw[fill=black, line width = 0.3mm, shift={(0,0)}]  (8.95,-1.036)circle(\kleinrad);
\draw[fill=none, line width = 0.1mm, shift={(0,0)}] (8.95,-1.036) circle(\myrad);
\draw[fill=black, line width = 0.3mm, shift={(0,0)}]  (5.67,14.833)circle(\kleinrad);
\draw[fill=none, line width = 0.1mm, shift={(0,0)}] (5.67,14.833) circle(\myrad);
\draw[fill=black, line width = 0.3mm, shift={(0,0)}]  (6.843,-3.483)circle(\kleinrad);
\draw[fill=none, line width = 0.1mm, shift={(0,0)}] (6.843,-3.483) circle(\myrad);
\draw[fill=black, line width = 0.3mm, shift={(0,0)}]  (-1.5,3.412)circle(\kleinrad);
\draw[fill=none, line width = 0.1mm, shift={(0,0)}] (-1.5,3.412) circle(\myrad);
\draw[fill=black, line width = 0.3mm, shift={(0,0)}]  (2.793,3.271)circle(\kleinrad);
\draw[fill=none, line width = 0.1mm, shift={(0,0)}] (2.793,3.271) circle(\myrad);
\draw[fill=none, line width = 0.1mm, shift={(0,0)}] (14.413,1.833) circle(\myrad);
\draw[fill=black, line width = 0.3mm, shift={(0,0)}]  (7.851,7.169)circle(\kleinrad);
\draw[fill=none, line width = 0.1mm, shift={(0,0)}] (7.851,7.169) circle(\myrad);
\draw[fill=black, line width = 0.3mm, shift={(0,0)}]  (-0.006,13.907)circle(\kleinrad);
\draw[fill=none, line width = 0.1mm, shift={(0,0)}] (-0.006,13.907) circle(\myrad);
\draw[fill=black, line width = 0.3mm, shift={(0,0)}]  (-5.232,-0.409)circle(\kleinrad);
\draw[fill=black, line width = 0.3mm, shift={(0,0)}]  (4.691,-2.111)circle(\kleinrad);
\draw[fill=none, line width = 0.1mm, shift={(0,0)}] (4.691,-2.111) circle(\myrad);
\draw[fill=black, line width = 0.3mm, shift={(0,0)}]  (10.407,7.908)circle(\kleinrad);
\draw[fill=none, line width = 0.1mm, shift={(0,0)}] (10.407,7.908) circle(\myrad);
\draw[fill=black, line width = 0.3mm, shift={(0,0)}]  (-0.479,0.724)circle(\kleinrad);
\draw[fill=none, line width = 0.1mm, shift={(0,0)}] (-0.479,0.724) circle(\myrad);
\draw[fill=black, line width = 0.3mm, shift={(0,0)}]  (-2.922,15.545)circle(\kleinrad);
\draw[fill=black, line width = 0.3mm, shift={(0,0)}]  (-2.868,8.662)circle(\kleinrad);
\draw[fill=none, line width = 0.1mm, shift={(0,0)}] (-2.868,8.662) circle(\myrad);
\draw[fill=black, line width = 0.3mm, shift={(0,0)}]  (5.19,1.444)circle(\kleinrad);
\draw[fill=none, line width = 0.1mm, shift={(0,0)}] (5.19,1.444) circle(\myrad);
\draw[fill=black, line width = 0.3mm, shift={(0,0)}]  (-2.705,-0.026)circle(\kleinrad);
\draw[fill=none, line width = 0.1mm, shift={(0,0)}] (-2.705,-0.026) circle(\myrad);
\draw[fill=black, line width = 0.3mm, shift={(0,0)}]  (1.225,-0.054)circle(\kleinrad);
\draw[fill=none, line width = 0.1mm, shift={(0,0)}] (1.225,-0.054) circle(\myrad);
\draw[fill=black, line width = 0.3mm, shift={(0,0)}]  (-2.718,5.792)circle(\kleinrad);
\draw[fill=none, line width = 0.1mm, shift={(0,0)}] (-2.718,5.792) circle(\myrad);
\draw[fill=black, line width = 0.3mm, shift={(0,0)}]  (6.259,2.829)circle(\kleinrad);
\draw[fill=none, line width = 0.1mm, shift={(0,0)}] (6.259,2.829) circle(\myrad);
\draw[fill=black, line width = 0.3mm, shift={(0,0)}]  (-4.702,2.614)circle(\kleinrad);
\draw[fill=black, line width = 0.3mm, shift={(0,0)}]  (7.828,12.07)circle(\kleinrad);
\draw[fill=none, line width = 0.1mm, shift={(0,0)}] (7.828,12.07) circle(\myrad);
\draw[fill=none, line width = 0.1mm, shift={(0,0)}] (14.878,3.326) circle(\myrad);
\draw[fill=black, line width = 0.3mm, shift={(0,0)}]  (3.435,9.172)circle(\kleinrad);
\draw[fill=none, line width = 0.1mm, shift={(0,0)}] (3.435,9.172) circle(\myrad);
\draw ( 6.8 , 2.9 ) node[anchor=center, scale=\csize, shift={(0,0)}]{$1$};
\draw ( 8.3 , 7.2 ) node[anchor=center, scale=\csize, shift={(0,0)}]{$2$};
\draw ( 2.9 , 9.15 ) node[anchor=center, scale=\csize, shift={(0,0)}]{$3$};
\draw ( 2.6 , 3.8 ) node[anchor=center, scale=\csize, shift={(0,0)}]{$4$};

\draw[color = black, fill=none, draw=black, line width = 0.3mm, shift={(0,0)}] (0,0) rectangle (10.0,10.0);
\end{tikzpicture}
}
\caption{Confetti of radius $r=0.4$ with Voronoi cells. The joint corner of the Voronoi cells 4, 1 and 2 has a distance larger than the radius of the circles to each of the center points of the cells 4, 1 and 2. Hence this corner is not covered by confetti.}
\label{F2}
\end{figure}
Checking if every point in \verb"S1" is covered requires some thought. While a discretization scheme would be possible, we felt that an approach via Voronoi cells is more elegant and in our preliminary studies produced less numerical errors. Recall that for a finite set of points $A\subset\R^2$ the Voronoi cells are defined as follows. For each $x\in A$, the Voronoi cell
\begin{equation}
\label{E06}
\mathrm{Vor}_A(x):=\big\{y\in\R^2:\,\|x-y\|=\min\{\|z-y\|:\,z\in A\}\big\}
\end{equation}
is the set of points $y$ that are closer (or equally close) to $x$ than to any other point in $A$. If we intersect the Voronoi cells of the points \verb"P" in \verb"List_of_points" with \verb"S2", we get cells that are bounded by finite polygons. A point $x$ in \verb"S1" is visible, if all vertices in the corresponding polygon have a distance to $x$ less than $r$. See Figure~\ref{F2}. Hence we need to compute a Voronoi tessellation of \verb"List_of_points" and intersect the cells with the square \verb"S2".

We used the statistics software \verb"R" and the package {\tt deldir} that computes Voronoi tessellations and the package {\tt polyclip} that computes the intersections of polygons.

In order to check if a given point $x$ is visible, we use a similar procedure. First we sort out all points of distance larger than $2r$. Then we compute the Voronoi tessellation of those points (except $x$) and intersect it with a regular $2^n$-gon around $x$ that approximates $B_r(x)$. There is a vertex of distance larger than $r$ to its cell center if and only if the $2^n$-gon has at least one visible point. By approximating $B_r(x)$ by the inner and the outer regular $2^n$-gon for increasing $n$ until the results coincide, we can check if $B_r(x)$ is visible.
\section{Simulation results}
Note that simulations with a large radius $r$ waste computing time since we simulate confetti in \verb"S2" but count only the confetti in \verb"S0". The ratio of areas is smaller for larger $r$. On the other hand, for small $r$, we need to simulate many layers of confetti before the whole square S1 is filled. Performing test runs, it seems that $r=0.05$ gives a good compromise.
\begin{figure}[ht]
\centerline{\includegraphics[height=8.0cm]{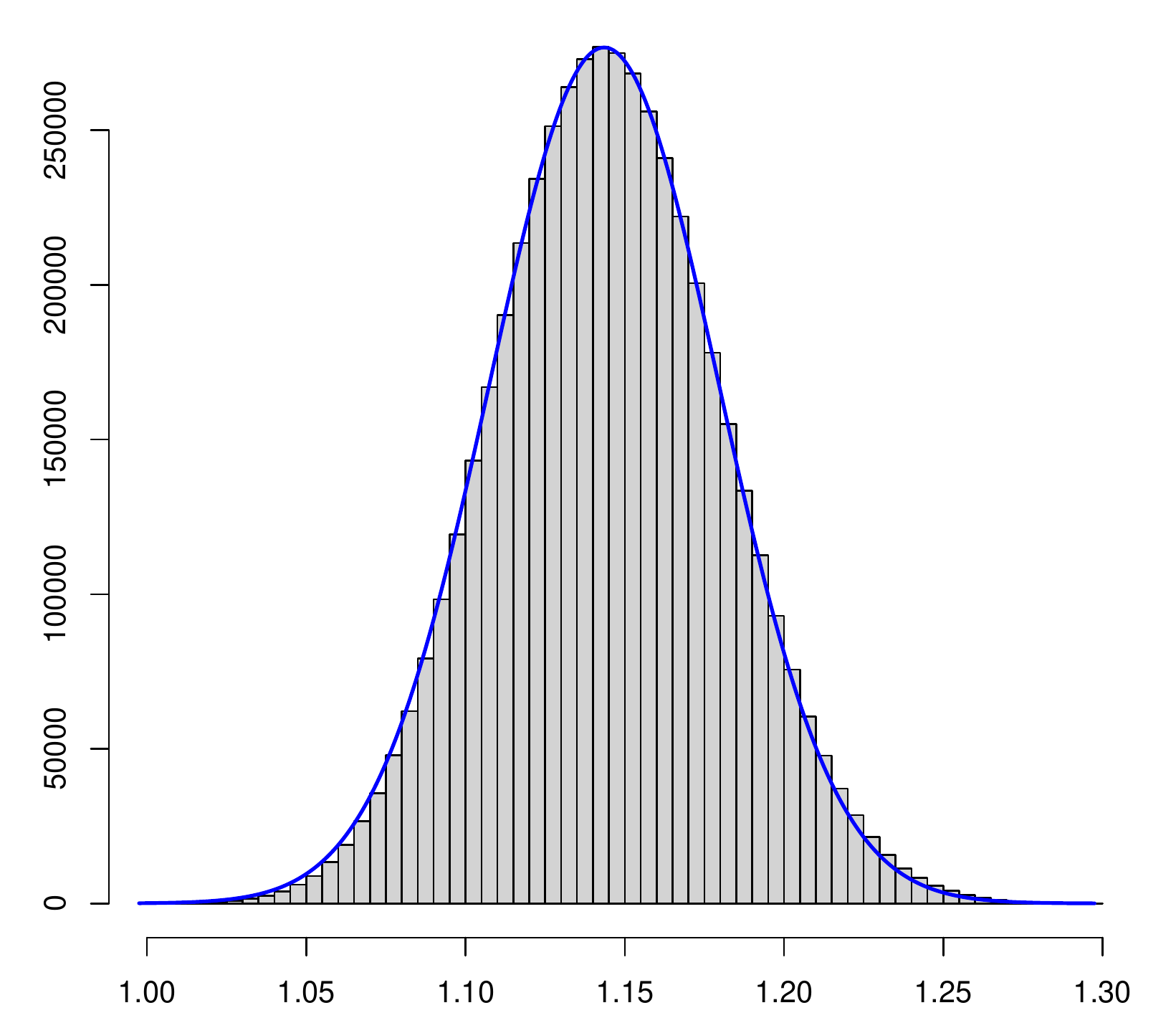}}
\caption{Histogram of the $C_i$ with normal density for comparison.}
\label{F3}
\end{figure}

We have performed a simulation with $r=0.05$ and sample size $n=5\,000\,000$. See \cite{KlenkeSimData2022} for a csv file of te simulation data. The simulations were performed at the \emph{Elwetritsch} high performance computing cluster at the university of Kaiserslautern. For each sample $i$, we have got a number $N_i$ of visible confetti in S0.
We let $C_i := 0.05^2N_i$.
Hence as an estimate for $c$, we get
\begin{equation}
\label{E07}
c\approx \bar C:=\frac{1}{n}\sum_{i=1}^nC_i=1.146015.
\end{equation}
The sample standard deviation for the $C_i$ is
\begin{equation}
\label{E08}
s:=\left(\frac{1}{n-1}\sum_{i=1}^n\big(C_i-\bar C\big)^2\right)^{1/2}=0.03603065
\end{equation}
Hence, the $2\sigma$ error for the estimate $\bar C$ is
\begin{equation}
\label{E09}
2s/\sqrt{5\,000\,000}=0.0000322.
\end{equation}

Clearly, since the correlations between the confetti are short range, we should expect a central limit theorem for the data $C_i$. The histogram supports this, see Figure~\ref{F3}.

\subsection*{Acknowledgments}
The author wishes to thank Jan Lukas Igelbrink who did a great job running the code. We also thank the \emph{Allianz für Hochleistungsrechnen Rheinland-Pfalz} for granting access to the High Performance Computing Cluster \emph{Elwetritsch} on which the simulations were run.

\def\cprime{$'$}

\end{document}